\documentclass[11pt,a4paper]{article}
\pdfoutput=1
\usepackage{atlasphysics}
\usepackage[displaymath]{lineno}
\usepackage{jheppub}
\usepackage[section]{placeins}
\usepackage{subfigure}
\usepackage{tabulary}
\usepackage{booktabs}
\usepackage{mathrsfs}
\usepackage{rotating}
\usepackage{float} 
\usepackage{url}
\usepackage{subfigure}
\usepackage{epstopdf}
\usepackage{rotating}
\usepackage{multirow}
\usepackage{xspace}
\usepackage{hyperref}
\hypersetup{
  colorlinks=true,
  linkcolor=red,
  citecolor=red,
  urlcolor=blue
}
\usepackage{preprintcover}  
\PreprintCoverPaperTitle{A search for $t \bar{t}$ resonances in lepton+jets events with highly boosted top quarks collected in $pp$ collisions at $\sqrt{s} = 7$~\tev{} with the ATLAS detector}  
\PreprintIdNumber{CERN-PH-EP-2012-158}  
\PreprintCoverAbstract{A search for resonant production of high-mass top-quark pairs is performed on 2.05~\ifb{} of proton-proton collisions at $ \sqrt{s} = 7$~\tev\ collected in 2011 with the ATLAS experiment at the Large Hadron Collider. This analysis of the lepton+jets final state is specifically designed for the particular topology that arises from the decay of highly boosted top quarks. 
The observed \ttbar{} invariant mass spectrum is found to be compatible with the Standard Model prediction and 95\% credibility level upper limits are derived on the \ttbar{} production rate through new massive states. An upper limit of 0.7 pb is set on the production cross section times branching fraction of a narrow 1~\tev{} resonance. A Kaluza--Klein gluon with a mass smaller than \upperKKgmasslimit~\tev{} is excluded.}  
\PreprintJournalName{Journal of High Energy Physics}  

\newcommand{\upperZprimemasslimit} {\ensuremath{1.15}}
\newcommand{\upperKKgmasslimit} {\ensuremath{1.5}}

\title{A search for $t \bar{t}$ resonances in lepton+jets events with highly boosted top quarks collected in $pp$ collisions at $\sqrt{s} = 7$~\tev{} with the ATLAS detector}

\author{The ATLAS Collaboration}

\date{version 1.4: \today}

\abstract{
A search for resonant production of high-mass top-quark pairs is performed on 2.05~\ifb{} of proton-proton collisions at $ \sqrt{s} = 7$~\tev\ collected in 2011 with the ATLAS experiment at the Large Hadron Collider. This analysis of the lepton+jets final state is specifically designed for the particular topology that arises from the decay of highly boosted top quarks. 
The observed \ttbar{} invariant mass spectrum is found to be compatible with the Standard Model prediction and 95\% credibility level upper limits are derived on the \ttbar{} production rate through new massive states. An upper limit of 0.7 pb is set on the production cross section times branching fraction of a narrow 1~\tev{} resonance. A Kaluza--Klein gluon with a mass smaller than \upperKKgmasslimit~\tev{} is excluded.
}

\begin{document}
\maketitle

\section{Introduction}

The Large Hadron Collider (LHC) opens up a new kinematic regime, where 
pairs of Standard Model (SM) particles can be produced with an invariant mass 
of several \tev. Such high-mass final states are of particular interest 
for searches for massive states predicted by a number of 
extensions of the Standard Model. 
High-mass pairs of top quarks are among the most interesting of the final states explored by the ATLAS~\cite{Aad:2008zzm} experiment, but also represent
a considerable experimental challenge.

The topology that forms when these Lorentz-boosted top quarks decay differs in important respects from that encountered when top quarks are produced approximately at rest. 
New tools have been developed to fully exploit the potential of these states. 
We adopt a solution proposed by Seymour~\cite{Seymour:1993mx} for the reconstruction and identification of highly boosted, hadronically decaying, 
massive particles, where these {\em boosted objects} are reconstructed as a single {\em fat} jet. 
An overview of the tools developed for the reconstruction of boosted objects is found in Refs.~\cite{boost2010, boost2011}. 

In this paper results are presented of a resonance search in the lepton+jets final state that arises in the reaction $pp \rightarrow t\bar{t} \rightarrow b\bar{b} q\bar{q}' \ell \nu_{\ell}$, 
where one of the $W$ bosons from the top quarks decays to a charged lepton and a neutrino, and the other to jets of hadrons. 
Events are classified as belonging to the ``$e$+jets'' or ``$\mu$+jets'' channel, depending on whether the charged lepton is an electron or a muon. 
We search for the distinct shape of a resonant signal in the reconstructed \ttbar{} invariant mass distribution.

Compared to searches for $t\bar{t}$ resonances carried out by the CDF~\cite{cdf_physrevlett_100,:2007dia,Aaltonen:2009tx,Aaltonen:2011ts,Aaltonen:2011vi} and D0~\cite{Abazov:2008ny,Abazov:2011gv} collaborations at Run II of the Fermilab Tevatron Collider and a previous search by ATLAS~\cite{resolved} using the present data set, 
this analysis is specifically designed for top-quark pairs with an invariant mass beyond 1~\tev{}~\cite{ATL-PHYS-PUB-2010-008}. 
Jets are reconstructed with the anti-$k_t$ algorithm~\cite{Cacciari:2008gp} with a larger radius parameter ($R= 1.0$) than is usually employed in ATLAS. 
The highly energetic top quark decaying to three jets of hadrons ($t \rightarrow Wb \rightarrow b q \bar{q}'$) is reconstructed as a single fat jet. 
The selection relies strongly on an analysis of the jet substructure. 
Also the reconstruction of the second top quark candidate (with the decay $t \rightarrow Wb \rightarrow b \ell \nu_{\ell}$) relies on the large boost of the top quarks; 
the jet assignment is based on the vicinity to the charged lepton originating from the top quark decay.

While \ttbar{} resonance searches are relevant for any extension of the Standard Model that leads to an enhanced top quark pair production rate at large \ttbar{} invariant mass, 
we interpret the result within two specific benchmark models. 
The leptophobic topcolor \Zprime{} boson\footnote{The specific case considered here corresponds to model IV in Ref.~\cite{topcolor2} with $f_1 =1$ and $f_2= 0$ and a width of 1.2\% of the \Zprime{} boson mass.}~\cite{topcolor2} represents an example of a narrow resonance, 
where the experimental resolution dominates the width of the reconstructed mass peak. 
The Tevatron searches have set a 95\% credibility level (CL) limit on the mass of the leptophobic topcolor \Zprime{} boson~\cite{Hill:1993hs} at $m_{Z'} > 900$~\geV~\cite{Aaltonen:2011ts}.  
The second benchmark model envisages a Kaluza--Klein (KK) excitation of the gluon $g_{KK}$, as predicted in models with warped extra dimensions~\cite{Agashe:2006hk,Lillie:2007ve}. 
For the choice of parameters of Lillie et al.~\cite{Lillie:2007yh} used here, 
the KK gluon manifests itself as a relatively broad resonance ($\Gamma/m = 15.3$\%) with a branching fraction $BR(g_{KK} \rightarrow t \bar{t}) = 92.5$\%. 
The first $t\bar{t}$ resonance searches on LHC data~\cite{resolved, cmsboosted} exclude Kaluza--Klein gluons~\cite{Agashe:2006hk,Lillie:2007ve} with a mass smaller than 1.13~\tev~\cite{resolved} at 95\% CL.

\section{The ATLAS detector}
The ATLAS detector~\cite{Aad:2008zzm} is a multi-purpose particle detector with a forward-backward symmetric cylindrical geometry and almost $4 \pi$ coverage in solid angle.

The inner detector (ID), composed of a silicon pixel detector, a silicon microstrip detector and a transition radiation tracker, provides efficient reconstruction of the trajectories of charged particles in the pseudorapidity\footnote{ATLAS uses a right-handed coordinate system with its origin at the nominal interaction point in the centre of the detector and the $z$-axis along the beampipe. The azimuthal angle $\phi$ is measured with respect to the $x$-axis, which points towards the centre of the LHC ring. The $y$-axis points up. The pseudorapidity $\eta$ is defined in terms of the polar angle $\theta$ as $\eta = - \ln{\tan{\theta/2}}$. The transverse momentum \pt\ is defined as $\pt = p \sin \theta$.} 
range up to $|\eta|=2.5$. 

The ID is surrounded by a thin superconducting solenoid producing a 2 T magnetic field and by a hermetic calorimeter system, which provides three-dimensional reconstruction of particle showers up to $|\eta| = 4.9$. 
A highly granular lead and liquid-argon (LAr) sampling calorimeter provides a precise measurement of the energy of electrons and photons. 
The hadronic calorimeter uses steel and scintillating tiles in the barrel region ($|\eta| <1.7$), while the endcaps use LAr as the active material and copper as absorber. 
The forward calorimeter ($|\eta| > 3.1$) also uses LAr as the active medium, with copper and tungsten as absorber. 

The muon spectrometer consists of one barrel and two
endcap air-core toroidal magnets, each consisting of eight superconducting
coils arranged symmetrically in azimuth around
the calorimeter. Three layers of precision tracking chambers, consisting 
of drift tubes and cathode strip chambers,
allow precise muon momentum measurement up to
$|\eta| =$ 2.7. Resistive plate and thin-gap chambers provide
muon triggering capability up to $|\eta| =$ 2.4.

The trigger system is composed of three consecutive
levels. The level-1 trigger is based on custom-built hardware that processes 
coarse detector information to reduce the event rate
to a design value of at most 75 $\mathrm{kHz}$. This is followed by two
software-based trigger levels, level-2 and the event filter, which
together reduce the event rate to a few hundred $\mathrm{Hz}$ which is 
recorded for analysis.

\section{Data and Monte Carlo samples}
\label{sec:samples}

The data used in this search were collected by the ATLAS detector at the LHC in 2011 using a single-muon or single-electron trigger with transverse momentum ($p_{\mathrm{T}}$) thresholds set at 18~\gev\ for muons and 20~\gev{} or 22~\gev\ for electrons. 
The object requirements used in the offline selection are more stringent than those used  in the trigger, and the offline $p_{\mathrm{T}}$ thresholds are chosen on the efficiency plateau for the trigger. 
Only data recorded under stable beam conditions between March and August 2011 are used. 
Moreover, all subdetectors are required to be operational. 
The resulting data sample corresponds to an integrated luminosity of \mbox{$2.05 \pm 0.08$~\ifb{}~\cite{Aad:2011dr, ATLAS-CONF-2011-116}}.  

Simulated samples are used to predict the contribution of the Standard Model backgrounds, the most important of which are \ttbar{} production, vector boson production in association with jets and multijet production. 
Monte Carlo (MC) simulations are also used to evaluate the impact of systematic uncertainties on the modelling of initial- and final-state radiation, the \ttbar{} production process, as well as on parton showering and hadronization. 

The irreducible ``continuum'' $t\bar{t}$ background and electroweak single top quark production are generated using {\sc MC@NLO v3.41}~\cite{FRI-0201,FRI-0301,Frixione:2005vw,Frixione:2008yi} with {\sc CTEQ6.6}~\cite{PhysRevD.78.013004} parton distribution functions (PDFs). 
Parton showering and hadronization are performed using {\sc Herwig v6.5}~\cite{COR-0001, Corcella:2002jc} in association with {\sc Jimmy}~\cite{jimmy} to model effects due to the underlying event and multiple parton interactions. 
The total production cross sections are based on approximate next-to-next-to-leading-order (NNLO) calculations. The pair production cross section is taken to be
165 $\mathrm{pb}$~\cite{Moch:2008qy,Beneke:2009ye,Aliev:2010zk}. 
For single top quark production, 65 $\mathrm{pb}$ ($t$-channel~\cite{Kidonakis:2011wy}), 4.6 $\mathrm{pb}$ ($s$-channel~\cite{Kidonakis:2010tc}) and 15.7 $\mathrm{pb}$ ($Wt$ associated production~\cite{Kidonakis:2010ux}) are used. 
Samples generated with {\sc AcerMC}~\cite{Kersevan:2004yg} and {\sc Powheg}~\cite{FRI-0701}, showered with either {\sc Herwig} or {\sc Pythia}~\cite{pythia}, are used to evaluate systematic uncertainties on the modelling of initial and final state radiation, the \ttbar{} production process, as well as on the parton showering and hadronization.

Vector boson production with associated jets ($V+$jets) is simulated with the {\sc Alpgen v2.13}~\cite{MAN-0301} generator with {\sc CTEQ6L1}~\cite{Pumplin:2002vw} PDFs. 
Only leptonic vector boson decays ($W \rightarrow \ell \nu_{\ell}$, $Z \rightarrow \ell^+ \ell^-$) are considered for these backgrounds.
Events are generated in exclusive bins of parton multiplicity up to four, and inclusively for larger multiplicity. 
The events are showered with {\sc Herwig} and {\sc Jimmy}. 
Matching of parton showers to the matrix elements avoids double counting of parton emissions in both parts of the calculation.
The normalization of the $W+$jets yield is derived from data, 
as described in Section~\ref{sec:backgroundwjets}.
The $Z+$jets sample, which has a much smaller contribution to the signal region,
 is normalized to the inclusive NNLO cross section~\cite{Hamberg:1990np}.

Diboson samples are produced using~{\sc Herwig v6.5} with {\sc MRST2007LO*}~\cite{Sherstnev:2007nd} PDFs and {\sc Jimmy}. 
A filter is applied at the generator level that requires the presence of one lepton with \pt\ $ > $  10 GeV and $|\eta| <$ 2.8.  
K-factors are applied such that the production cross sections agree with results obtained using the MC@NLO Monte Carlo generator and the MSTW2008 PDF set~\cite{Martin:2009iq}.
The cross sections (K-factors) used for these filtered samples are: 11.5  $\mathrm{pb}$ (1.48) for $WW$ production, 3.46 $\mathrm{pb}$ (1.60) for $WZ$ production, and 0.97 $\mathrm{pb}$ (1.30) for $ZZ$ production.

Signal samples for the $pp \rightarrow \Zprime \rightarrow t \bar{t}$ process
are generated using {\sc Pythia v6.421} with {\sc CTEQ6L1} PDFs. Kaluza--Klein 
gluons are generated with {\sc Madgraph v4.4.51}~\cite{madgraph,Alwall:2007st} 
with {\sc CTEQ6L1} PDFs, and showered with {\sc Pythia}. 
Interference with Standard Model \ttbar{} production is not taken into account.
The production cross sections times branching fraction used for both benchmark signal models are presented in Table~\ref{tab:signalxsec}. 
The production cross sections for the \Zprime{} boson samples are evaluated as in Ref.~\cite{Harris:2011ez} and a K-factor of 1.3 is applied to account for next-to-leading-order (NLO) effects~\cite{Gao:2010bb}. 
The KK gluon production cross sections are determined using {\sc Pythia} v8.1~\cite{Sjostrand:2007gs}.

\begin{table}[tbh]
\begin{center}
\caption{Production cross sections times branching fraction for the resonant signal processes $pp \rightarrow \Zprime{} \rightarrow t \bar{t}$ in the topcolor model and $pp \rightarrow g_{KK} \rightarrow t \bar{t}$ for the KK gluon in Randall--Sundrum models with warped extra dimensions.}
\vspace{2mm}
\begin{tabular}{lcccccccccc}
\hline
Mass [\gev{}] & 600 & 700 & 800 & 900 & 1000 & 1200 & 1400 & 1600 & 1800 & 2000   \\ \hline
$\sigma_{Z'} \times$BR [$\mathrm{pb}$]  & 10.3 & 5.6 & 3.2 & 1.9 & 1.2 & 0.46 & 0.19 & 0.068 & 0.039 & 0.018  \\ 
 $\sigma_{g_{KK}} \times$BR [$\mathrm{pb}$]  & 39.4 & 20.8 & 11.6 & 6.8 & 4.1 & 1.7 & 0.73 & 0.35 & 0.18 & 0.095\\ \hline
\end{tabular}
\label{tab:signalxsec}
\end{center}
\end{table} 

All generated samples 
are processed using a {\sc GEANT4}-based~\cite{AGO-0301} simulation of the 
ATLAS detector~\cite{:2010wqa} and reconstructed with the
reconstruction software used also for data. The trigger response is emulated 
in the offline software. 
A varying number of simulated minimum-bias events are overlaid on the hard process to account for the effect of multiple $pp$ interactions per bunch-crossing, which are quantified by the variable $\mu$. 
Then the simulated events are reweighted so that the data and the simulated sample have the same $\mu$ distribution.

\section{Physics object reconstruction and selection}
\label{sec:objects}

Electron candidates must have an electromagnetic (EM) shower shape consistent with expectations based on simulation, test beam and reconstructed $Z \rightarrow ee$ events in data, 
and must have a matching track in the ID~\cite{Aad:2011mk}.  
The EM cluster must be within \mbox{$|\eta_{\mathrm{cluster}}| < 2.47$}, excluding the calorimeter transition region at \mbox{$1.37 < |\eta_{\mathrm{cluster}}| < 1.52$}. 
The isolation transverse energy is determined as the transverse component of the sum of the energy deposits found in the calorimeter in an $ \eta - \phi$ cone of radius $\Delta R = \sqrt{( \Delta \phi )^2 + ( \Delta \eta )^2}= 0.2$  around the electron position. 
The energy of the electron is subtracted and the energy deposited by particles from additional $pp$ interactions is accounted for by applying a correction that depends on the number of primary vertices. 
The contamination by non-isolated electrons due to decays of hadrons (including heavy flavour) in jets is reduced by requiring the corrected isolation transverse energy to be less than 3.5~\gev.

Muon candidates are reconstructed from track segments in different layers of the muon chambers. 
These segments are then combined, starting from the outermost layer, with a procedure that takes material effects into account, and matched with tracks found in the inner detector. 
The candidate trajectories are  refitted using the complete track information from both detector systems, and are required to satisfy $| \eta | < 2.5$. 
Non-isolated muons are rejected by requiring that the energy deposited in the calorimeter and the scalar sum of track transverse momenta in a cone of $ \Delta R =0.3$ around the muon candidate, after subtraction of the muon energy deposit or $p_{\mathrm{T}}$, are both less than 4~\gev. 

The isolation requirement on the leptons results in some loss of signal efficiency at high \ttbar\ masses, as shown in Section~\ref{sec:selection}.

Calorimeter cells are clustered using a three-dimensional representation of the energy depositions in the calorimeter with a nearest-neighbour noise-suppression algorithm~\cite{TopoClusters, Aad:2011he}. 
Such topological clusters form the input to the jet reconstruction algorithm. 
Two types of jets are used, both reconstructed with the anti-$k_t$ algorithm~\cite{Cacciari:2008gp}. 
For the first type, a radius parameter $R=0.4$ is used. 
The input to the jet algorithm is formed by topological clusters calibrated at the EM energy scale, 
appropriate for the energy deposited by electrons or photons. 
A second set of jets is created with a radius parameter $R=1.0$. 
These jets are henceforth referred to as fat jets. 
The input to this second jet reconstruction is formed by locally calibrated topological clusters~\cite{ISS-0501}. 
In the {\em local} calibration, clusters are classified as hadronic or electromagnetic based on the cluster shape, depth and energy density. 
A correction is applied to the cluster that depends on this classification. 
Locally calibrated topological clusters are thus corrected for calorimeter non-compensation and are typically within 10\% of the energy scale of the corresponding particles. 
In both cases the jet transverse momentum and pseudorapidity are corrected using $p_{\mathrm{T}}$- and $\eta$- dependent calibration factors obtained from simulation~\cite{Aad:2011he} and validated with collision data~\cite{Aad:2011he,Aad:2012ef}. 
The validity of the calibration of fat jets is limited to absolute rapidity smaller than 2 and jets outside this region are not considered. 
For the fat jets a further correction is applied to the jet invariant mass\footnote{The jet four-vector is obtained by summing the four-vectors of the (massless) clusters in the calorimeter associated with the jets. As the jet components have non-zero opening angle, even jets that result from the hadronization of massless gluons or light quarks acquire a non-zero mass~\cite{Aad:2012ef}.}.

The magnitude \MET{} of the missing transverse momentum is constructed from the vector sum of the energy deposits in calorimeter cells associated with topological clusters. 
Calorimeter cells are associated with a parent physics object in a chosen order: electrons, jets reconstructed with $R=0.4$, and muons, such that a cell is uniquely associated with a single physics object. 
Cells belonging to electrons are calibrated at the electron energy scale, and double counting of cell-energies is avoided, while cells belonging to jets are taken at the corrected energy scale used for jets. 
Finally, the track $p_{\mathrm{T}}$ of muons passing the event selection is included, 
and the contributions from any calorimeter cells associated with the muons are subtracted. 
The remaining clusters not associated with electrons or jets are included at the EM energy scale.

Overlap between the different object categories is avoided by the following procedure. 
Jets within $\Delta R = 0.2$ of an electron passing the  electron selection cuts are removed from the jet collection. Muons within $\Delta R =0.4$ of any $R= 0.4$ jet with \mbox{$p_{\mathrm{T}} >20$~\gev} are rejected. 
Subsequently, events where the selected electron is separated by less than $ \Delta R = 0.4$ of any jet reconstructed with $R= 0.4$ and with \mbox{$p_{\mathrm{T}} > 20$~\gev} are rejected. 

For all reconstructed objects in the simulation, scale factors and additional smearing are applied to compensate for the difference in reconstruction efficiencies between data and simulation. 
The uncertainties on these scale factors are used to determine the corresponding systematic uncertainties.

\section{Event selection and reconstruction of the \ttbar\ system}
\label{sec:selection}

Events accepted by the single-electron or single-muon trigger are required 
to have at least one reconstructed primary vertex with 
at least five associated tracks with $p_{\mathrm{T}} > 400$~\mev. 
Events are discarded if any jet with $p_{\mathrm{T}} >20$~\gev{} is identified as out-of-time activity, calorimeter noise, or is
located in a problematic area of the calorimeter~\cite{Aad:2011he}.

A single isolated charged lepton that meets the quality criteria of 
Section~\ref{sec:objects} is required. 
Selected electrons must match the online lepton candidate responsible for the trigger decision. 
Events where an electron shares an
inner detector track with a non-isolated muon are discarded. Muons are required 
to have a transverse momentum greater than 20~\gev{} and electrons must 
have $E_{\mathrm{T}} > 25$~\gev. 

\begin{figure}[htbp]
\begin{center}
\includegraphics[width=0.9\linewidth]{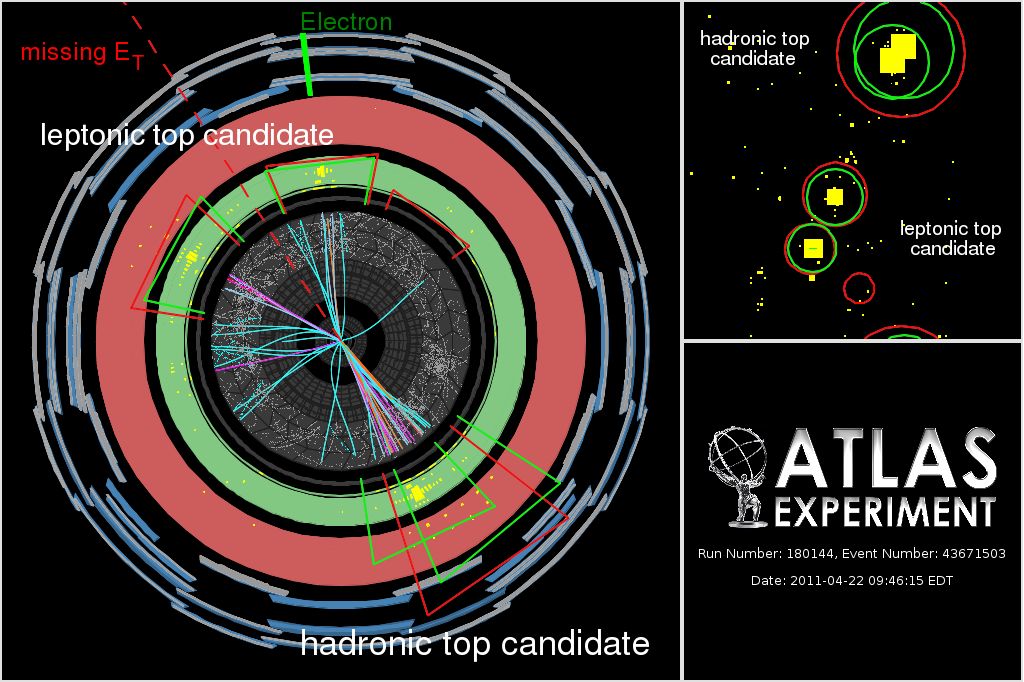}
\caption{Event display for a $t \bar{t}$ candidate event with large $t \bar{t}$
invariant mass: $m_{t\bar{t}}=2.5$~\tev. The left panel displays a transverse view of the charged particle tracks and calorimeter energy deposits. An $\eta-\phi$ view of the same event is shown in the upper right panel. Jets reconstructed with $R= 0.4$ are indicated in green, jets with $R= 1.0$ in red (colour online). \label{fig:display_electron}}
\end{center}
\end{figure}

The escaping neutrino from the leptonic $W$ boson decay leaves a signature in 
the $p_{\mathrm{T}}$ balance of the event. 
Different selections, optimized to suppress multijet events, are applied in the two channels.
In the $e$+jets channel, 
the magnitude of the missing transverse momentum $E_{\mathrm{T}}^{\mathrm{miss}}$ must be 
larger than 35~\gev{} and the transverse 
mass\footnote{The transverse mass is defined as $m_{\mathrm{T}} = \sqrt{2 p_{\mathrm{T}}^{\ell} E_{\mathrm{T}}^{\mathrm{miss}} ( 1 - \cos{\Delta \phi})}$, where
$p_{\mathrm{T}}^{\ell}$ is the charged lepton $p_{\mathrm{T}}$ and 
$\Delta \phi$ is the azimuthal angle between the lepton and the missing transverse momentum.} \mbox{$m_{\mathrm{T}}(\ell,E_{\mathrm{T}}^{\mathrm{miss}}) > 25$~\gev}. 
In the $\mu$+jets channel, \mbox{$E_{\mathrm{T}}^{\mathrm{miss}}> 20$~\gev\ }
is required, as well as \mbox{$E_{\mathrm{T}}^{\mathrm{miss}} + m_{\mathrm{T}}(\ell,E_{\mathrm{T}}^{\mathrm{miss}}) > 60$~\gev}.

Assuming the missing transverse momentum is dominated by the escaping neutrino from the $W$ boson decay, the neutrino
momentum can be reconstructed by imposing a $W$ mass constraint on the 
lepton-\MET{} system. If the quadratic equation yields two real solutions, 
the solution with the smallest $|p_z|$ is chosen. If the discriminant of the
quadratic equation is negative, the magnitude of the missing transverse energy 
is adjusted to get a null discriminant.
The selection steps based on the reconstructed charged lepton and the signature
of the escaping neutrino, and the reconstruction of the
leptonic $W$ boson candidate follow closely that of a previous analysis
of the same final state~\cite{resolved}.

The selection of jets and their assignment to top quark candidates is designed specifically for the collimated topology of the decay products of highly boosted top quarks. 
The lepton and jet from the semi-leptonically decaying top quark ($t \rightarrow W b \rightarrow b \ell \nu_{\ell}$) are expected to be collimated in a relatively small area of the detector.  
A search region with $0.4 < \Delta R(l,j) < 1.5$ is defined around the direction of the charged lepton. 
Events are accepted if at least one jet with $p_{\mathrm{T}} > 30$~\gev{} is found in this region. 
If several jets with $p_{\mathrm{T}} >30$~\gev{} are found, the one with smallest angular distance $\Delta R(\ell,j)$ to the lepton is retained. 
The semi-leptonic top candidate is then constructed by adding the four-momenta of the reconstructed lepton, the neutrino candidate and the selected jet.  

The decay products from a highly-boosted hadronically decaying top quark form a single fat jet that is expected to be found approximately {\em back-to-back} in $\phi$ to the semi-leptonic top decay (i.e. the two top quarks are emitted in opposite directions in the transverse plane, and the top quark boost ensures that their decay products retain the approximate direction of the top quarks). 
We require at least one ($R= 1.0$) jet at a minimum distance $ \Delta R(j,j) >  1.5$ from the jet associated with the semi-leptonic top candidate. 
With this $ \Delta R(j,j)$ requirement, moreover, we avoid that clusters of energy deposits in the calorimeter are shared between the two types of jets. 
The fat jet is required to have $p_{\mathrm{T}} > 250$~\gev{}. 
The fat jet mass $m_{j}$ is expected to reflect the large top quark mass, and is required to be greater than 100~\gev. 
Finally, the jet components are reclustered using the $k_t$ algorithm in {\sc FastJet}~\cite{Cacciari:2011ma}, and the last splitting scale $\sqrt{d_{12}}$~\cite{Aad:2012ef} is required to be greater than 40~\gev. 
If more than one fat jet is found, the leading $p_{\mathrm{T}}$ jet is retained as the hadronic top candidate. 
The four-vector momentum associated with the $t\bar{t}$ system is reconstructed by adding the four-momenta of the semi-leptonically decaying top quark candidate and the hadronically decaying top quark candidate.

\begin{figure}[htbp]
\begin{center}
\includegraphics[width=0.7\textwidth]{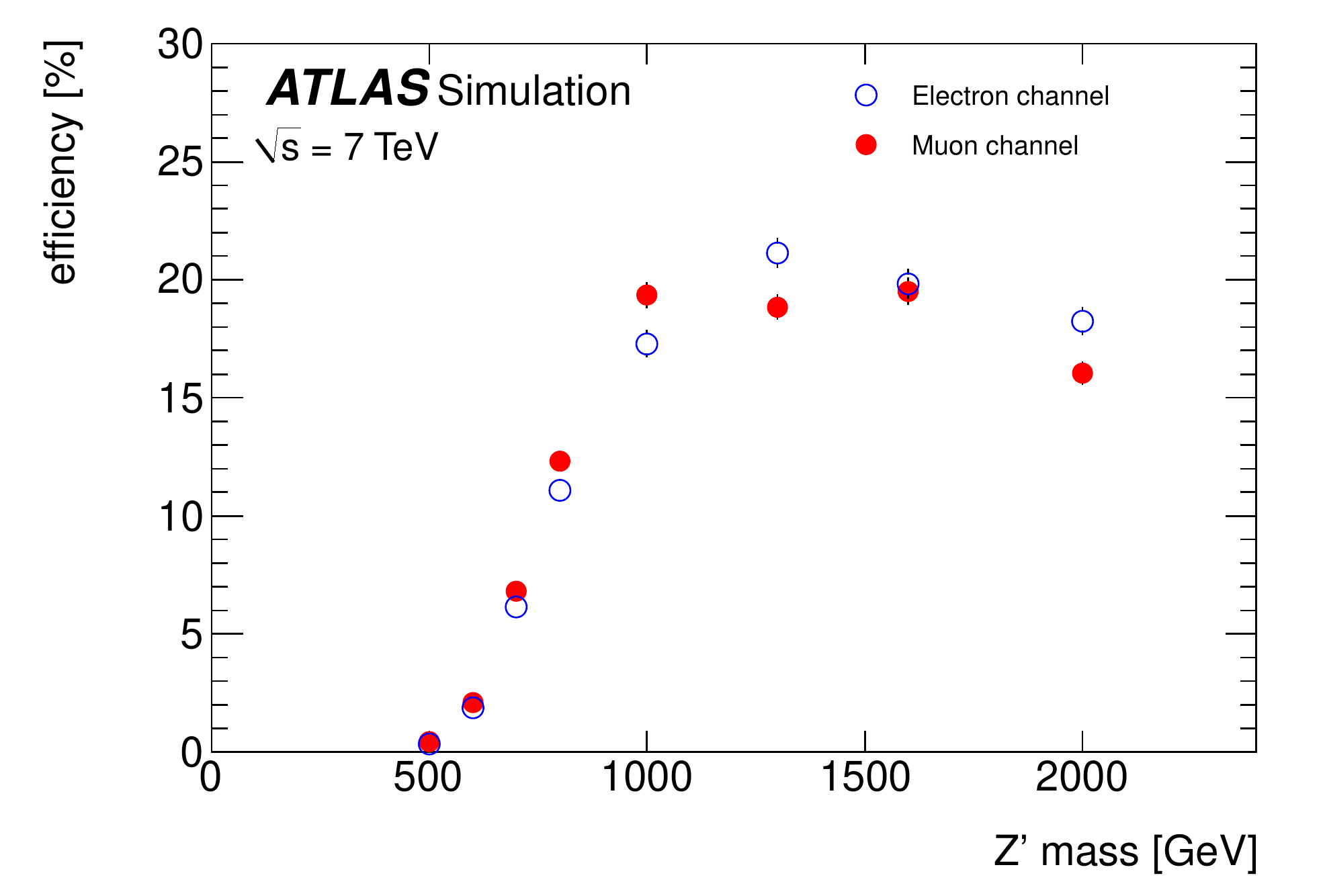}
\caption{Estimate from Monte Carlo simulation of the selection efficiency
for the leptophobic \Zprime{} benchmark model. 
Only events with the targeted final state are considered ($t \bar{t} \rightarrow W^+bW^{-}\bar{b} \rightarrow \ell \nu_{\ell} b \bar{b} j j$, where $\ell$ is either an electron or a muon, corresponding to approximately 30\% of \ttbar{} events). The error bars shown correspond to the Monte Carlo statistical uncertainty.
\label{fig:seleff_vs_zpmass}}
\end{center}
\end{figure}

\begin{figure}[htbp]
\begin{center}
\includegraphics[width=0.7\textwidth]{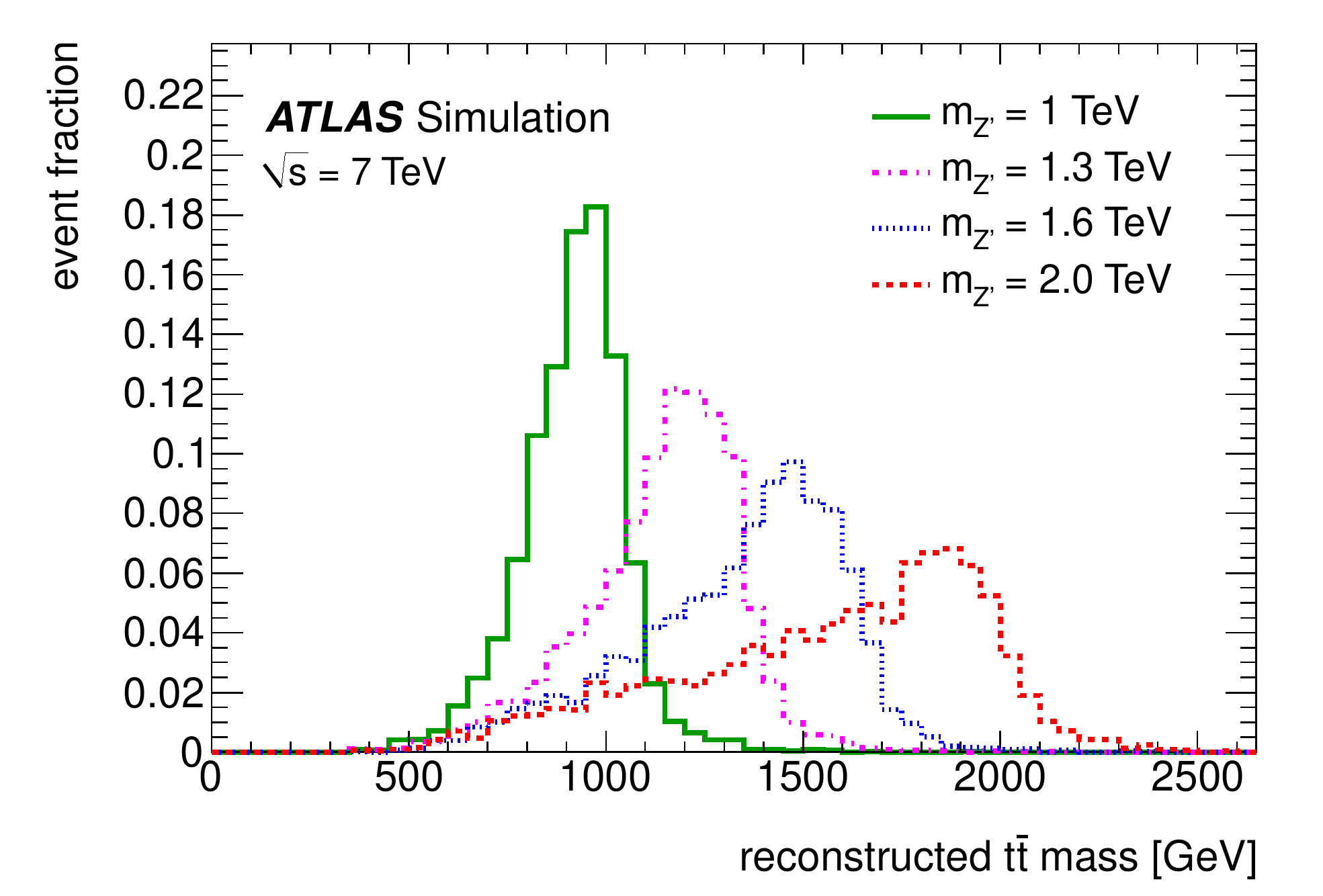}
\caption{Estimate from Monte Carlo simulation of the reconstructed  $t \bar{t}$ invariant mass distribution for the leptophobic \Zprime{}. 
The natural width of the resonance is small ($\Gamma/m = 1.2$\%) compared to the experimental mass resolution. 
For the highest mass points a considerable tail towards smaller mass appears due to the convolution with the rapidly falling parton luminosity.
\label{fig:mass_vs_zpmass}}
\end{center}
\end{figure}

The sample event display in Figure~\ref{fig:display_electron} illustrates the reconstruction procedure. 
Of the three jets reconstructed with $R= 0.4$, the jet closest to the lepton is associated with the leptonic top quark candidate. 
The two remaining $R= 0.4$ jets in the opposite hemisphere merge into a single fat jet ($p_{\mathrm{T}}= 641$~\gev{}, jet mass $m_{j} = 138$~\gev{}, $\sqrt{d_{12}}= 107$~\gev{}) when the event is reclustered with $R= 1.0$. 
The invariant mass of the system formed by the two top quark candidates is approximately 2.5~\tev{}.

The efficiency times acceptance for signal events due to resonant 
$pp \rightarrow \Zprime \rightarrow t\bar{t}$ production is shown 
in Figure~\ref{fig:seleff_vs_zpmass} and is seen to depend strongly on the 
resonance mass, with a relatively steep turn-on at approximately 800~\gev{}. 
For resonances with a mass greater than 2~\tev{} the acceptance is degraded, primarily due to the lepton isolation and the requirement on the minimum distance between the lepton and the nearest jet, because the top quarks are so highly boosted.

After this selection, the dominant background is Standard Model top-quark pair production. 
Two further Standard Model processes are expected to have an important contribution. 
The yield due to multijet production, where the isolated lepton signature is faked by leptons from heavy-flavour decays, muons from $\pi^{\pm}$ and $K^{\pm}$ decays in flight, electrons from photon conversions, or misidentified hadrons, is estimated in Section~\ref{sec:backgroundqcd}. 
The contribution of $W$ boson production in association with jets is normalized using data, as reported in Section~\ref{sec:backgroundwjets}. 
Further contributions from Standard Model processes are expected to be small, and their prediction is based on Monte Carlo estimates.

The reconstructed \ttbar{} mass distribution for four benchmark samples is shown in Figure~\ref{fig:mass_vs_zpmass}. 
For resonance masses between 1~\tev{} and 1.6~\tev{} the mass resolution, from a Gaussian fit to the distribution of reconstructed $t \bar{t}$ mass minus the true resonance mass, is approximately \mbox{10\%}. 
The tail that extends to lower masses is present to a lesser degree also in the generated \ttbar{} mass distribution, as a result of the convolution of the steeply falling parton luminosity with the resonance line shape. 
This effect is especially pronounced for the 1.6~\tev{} and 2~\tev{} points.

The reduced high-mass tail compared to Reference~\cite{resolved}
shows that this approach 
is intrinsically robust against the confusion that arises from the 
presence of additional jets due to initial-state 
radiation~\cite{ATL-PHYS-PUB-2010-008}. Thus, migration from the
low-mass region into the high-mass region is minimal.

\section{Estimate of the multijet background from data}
\label{sec:backgroundqcd}

To avoid the large systematic and statistical uncertainties in the MC prediction of events from multijet production, 
the contribution to the signal region is estimated from data using the Matrix Method~\cite{Aad:2012qf}. 
It exploits control regions with leptons satisfying looser identification requirements to disentangle the contributions from multijet production, 
which yields non-prompt leptons and jets misidentified as leptons, 
and sources of prompt leptons such as the production of a vector boson plus jets. 

The {\em loose} electrons have less stringent quality criteria. 
The electron isolation requirement is also modified: 
the total energy in a cone of $\Delta R =0.2$ around the electron is required to be smaller than 6~\gev{} (instead of 3.5~\gev{} for standard {\em tight} electrons), after correcting for energy deposits from event pile-up interactions and for the energy associated with the electron. 
The loose muon definition requires candidates to satisfy all criteria applied in the standard, tight selection except the muon isolation requirements and the muon-jet overlap removal.

The total number of events with loose leptons, $N_L$, is defined as
\begin{linenomath*}
\begin{equation}
N_L=N_{\mathrm{prompt}}+N_{\mathrm{multijets}}
\end{equation}
\end{linenomath*}
The subset of events with tight leptons should satisfy
\begin{linenomath*}
\begin{equation}
N_T=\epsilon{}\times{}N_{\mathrm{prompt}}+f\times{}N_{\mathrm{multijets}},
\end{equation}
\end{linenomath*}
where $\epsilon$ ($f$) indicates the probability that a prompt (multijet) lepton with loose selection criteria passes the standard, tight selection. 
Solving these two equations for $N_{\mathrm{prompt}}$ and $N_{\mathrm{multijets}}$, we estimate the multijets contribution to the signal region as:
\begin{linenomath*}
\begin{equation}
f\times{}N_{\mathrm{multijets}}=\frac{(\epsilon-1)\;f}{\epsilon-f}N_T+\frac{\epsilon\;f}{\epsilon-f}N_A,
\label{eqn:QCDMM}
\end{equation}
\end{linenomath*}
where $N_T$ is the number of events with a tight lepton, and $N_A$ is the number of events with a loose lepton which fail the tighter cuts of the standard selection. 
The fake rate $f$ is measured on a control sample rich in multijet events. 
The contamination of the control region by prompt leptons is estimated from MC simulation. The systematic uncertainties on $f$ include contributions from the uncertainty in the yield of the subtracted background sources and from the differences in the definition of the signal and multijet control regions. 

The probability $\epsilon$ is estimated using a tag-and-probe technique on a sample enriched in $Z\rightarrow \ell\ell$. 
Events with exactly two loose leptons of the same flavour and opposite electric charge are selected. 
At least one of the leptons must satisfy the tight quality criteria and the invariant mass of the two leptons is required to be between 86~\gev{}  and 96~\gev{}. 
The systematic uncertainties on $\epsilon$ cover the differences between the efficiency for $Z\rightarrow ee$ events and other sources of electrons, i.e. the background and signal processes considered in this analysis.

Kinematic distributions for the multijet background are constructed by assigning a weight to the events according to Equation~\ref{eqn:QCDMM}, with $(N_T,N_A)=(1, 0)$ for events with a tight lepton or $(0, 1)$ with a loose lepton that fails the tight cuts. 
$f$ and $\epsilon$ are largely independent of the kinematic variables of interest. Variations of the cuts on jet \pt, mass or split scale have no impact on $f$ within the statistical and systematic uncertainties. The efficiency $\epsilon$ is found to be constant for the invariant mass of the top quark candidates.

\section{$W$+jets production normalization and control regions}
\label{sec:backgroundwjets}

The $W$+jets contribution to the signal region is predicted using 
the {\sc Alpgen} sample described in Section~\ref{sec:samples}. To reduce
the uncertainty on the normalization, the total contribution to the signal region is estimated in situ using the observed charge asymmetry in 
data and the ratio in MC $r_{\mathrm{MC}}$ of $W$+jets events with positive leptons to those with negative leptons~\cite{Aad:2012ug}:
\begin{linenomath*}
\begin{equation}
(N_{W^+} + N_{W^-})^{\mathrm{pred}} = \left(\frac{r_{\mathrm{MC}} + 1}{r_{\mathrm{MC}} - 1}\right)(N_{W^+} - N_{W^-})^{\mathrm{data}}, 
\label{eqn:chargeasymmetry}
\end{equation}
\end{linenomath*}
where $N_{W^{+(-)}}$ is the number of predicted or
observed events with a positive (negative) lepton. The method is applied to 
a control region without jet mass and splitting scale requirements and a 
looser ($p_{\mathrm{T}} >150$ \gev) cut on the transverse
momentum of the hadronically decaying top quark candidate. All processes other than 
$W$+jets and SM \ttbar\ production have small contributions
 to this region and are subtracted. The fraction of $W$+jets events in the 
sample is determined using Equation~\ref{eqn:chargeasymmetry}. The resulting
scale factors, 0.77 (0.75) for the $e+$jets ($\mu+$jets) channel, are 
compatible with unity within the uncertainty. 
In the remainder of this paper, the $W$+jets MC prediction is scaled by the factors derived here. 
The uncertainty in the normalization and shape of the $W$+jets background are discussed in Section~\ref{sec:systematics}.

\begin{figure}[htb!]
\begin{center}
 \includegraphics[width=0.9\textwidth]{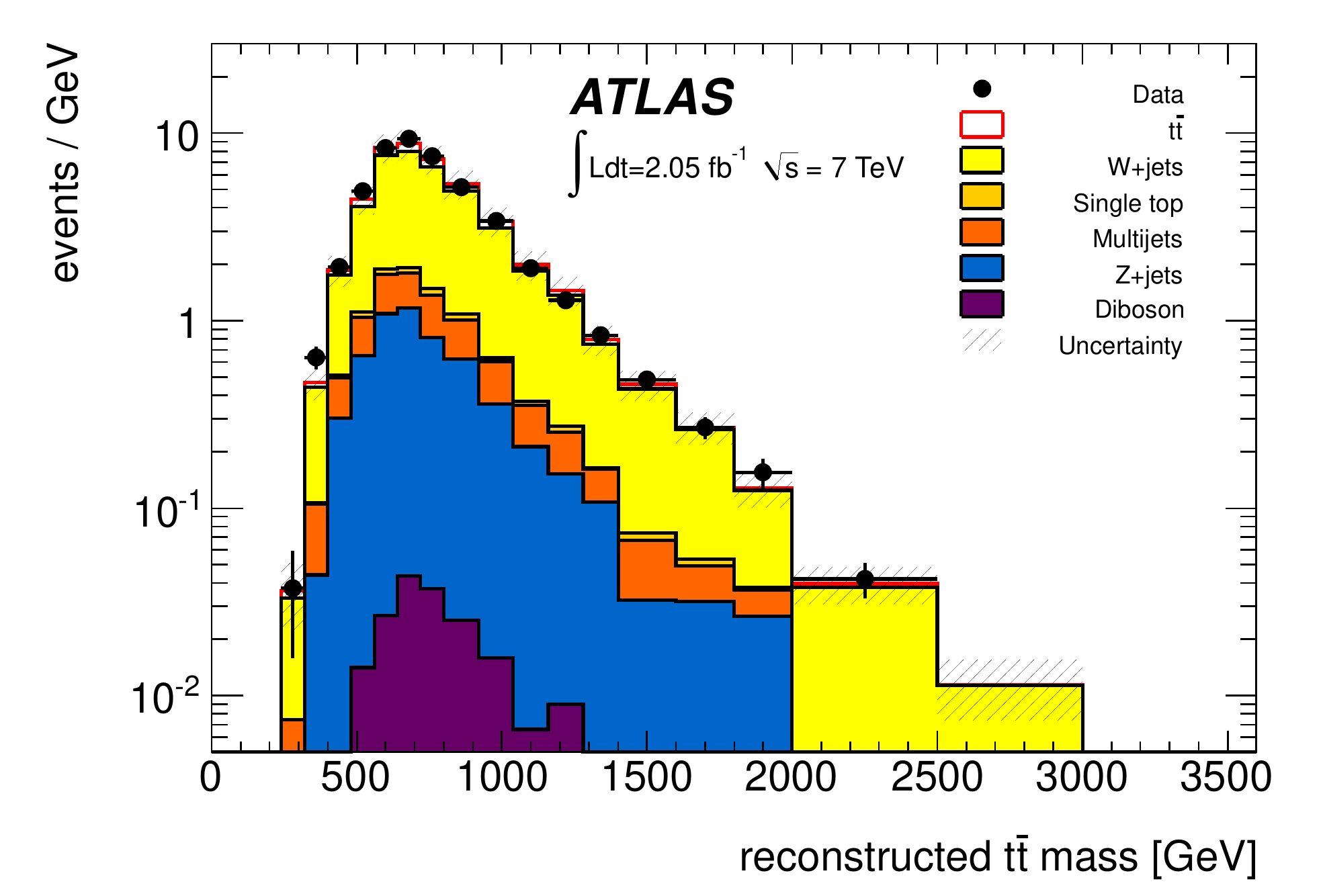}
\caption{The reconstructed $t \bar{t}$ invariant mass distribution of candidate events in the $W$+jets enriched control region. The $e$+jets and $\mu$+jets channels are combined. The scale factors derived in this section have been applied to the $W$+jets Monte Carlo distribution. The hatched area indicates the normalization uncertainty on the Standard Model prediction, but the shape uncertainty is not included. }
\label{fig:wjetsshapedata}
\end{center}
\end{figure}

As a cross-check, a second $W$+jets enriched control region is constructed, which differs from the signal region in that the jet mass and splitting scale requirements are removed. 
A $b$-jet veto is furthermore applied to reduce contamination due to \ttbar{} production, either from SM production or hypothetical sources beyond the Standard Model. 
The veto requires that a multivariate $b$-tagging algorithm~\cite{btagging} operated at a nominal $b$-tagging efficiency of 60\% in simulated $t\bar{t}$ events yields no positive tags for the jets reconstructed with $R= 0.4$. 
The observed invariant mass spectrum is compared to the SM prediction in Figure~\ref{fig:wjetsshapedata}. 
The $W$+jets contribution to the control region is over 70\%. 
After application of the data-driven scale factor, the observed and predicted number of events agree to within 3\% for both $e$+jets and $\mu$+jets channels, well within the uncertainty. 
The shape predicted by Monte Carlo is in good agreement with the observed distribution.

\section{Systematic uncertainties}
\label{sec:systematics}

A total of 30 sources of systematic uncertainty are taken into account. 
An overview is presented in Table~\ref{tab:systematics}. 
Systematic uncertainties with negligible impact on the sensitivity are omitted from the table, even if they are taken into account in the interpretation of the result. 
Groups of related uncertainties are combined. 
The relative impacts on the total expected background yield and signal yield are given for each source. 
Most uncertainties affect both the yield and the shape of the reconstructed mass distribution. 
 To estimate the impact of shape uncertainties, the effect on the limit on the production rate of a 1.3~\tev{} \Zprime{} boson is presented. 

\begin{table}[tbh!]
\begin{center}
\caption{Systematic uncertainties and their impact on the sensitivity. All uncertainties except ``luminosity'' and those labelled ``normalization'' affect the yield and the shape of the reconstructed mass distribution. In the first two columns the relative impact (in percent) is shown on the total expected background yield (nominally 1840 events) and on the number of selected signal events (a \Zprime{} with a mass 1.3~\TeV{} is chosen as the benchmark). 
The shape variations do not affect the overall normalization.  
The third column lists the relative variation for this benchmark of the expected limit on the production cross section times branching fraction if the corresponding systematic effect is ignored. 
The limit-setting procedure is explained in detail in Section~\ref{sec:limits}.
}
\vspace{2mm}
\normalfont{
\begin{tabular}{lccc}

\hline
Systematic effect & \multicolumn{2}{c}{~~Impact on yield [\%]} & \multicolumn{1}{c}{Impact on } \\
&background~~~&\Zprime 1.3 \TeV   & sensitivity [\%]\\
\hline
Luminosity                             &2.5 &3.7 & 0.4 \\
PDF uncertainty                        &3.1 &1.0 & 0.2\\
\ttbar\ normalization                  &4.9 & --- &  0.7\\
\ttbar\ ISR, FSR                       &6.3 & --- &  0.7\\
\ttbar\ fragmentation \& parton shower &3.4 & --- &  0.9\\
\ttbar\ generator dependence           &2.8 & --- &  2.2\\
$W$+ jets normalization                &4.3 & --- &  1.4\\
$W$+ jets shape                        &$norm.$& --- &  0.1\\
Multijets normalization                &2.1 & --- &  0.2\\
Multijets shape                        &$norm.$& --- &  1.1\\
$Z$+ jets normalization                &2.0 & --- &  0.5\\
Jet energy and mass scale              &6.7 &2.0 &  5.2\\
Jet energy and mass resolution         &4.7 &4.0 &  1.2\\
Electron ID and reconstruction         &1.1 &1.3 &  1.0\\
Muon ID and reconstruction             &2.2 &2.1 &  4.8\\
\hline
\end{tabular}
}
\label{tab:systematics}
\end{center}
\end{table}

Normalization and shape uncertainties on the most important Standard Model sources are estimated using a combination of in situ and Monte Carlo techniques. 
Following Ref.~\cite{Aad:2012qf}, three different systematic uncertainties take into account the imperfections in the modelling of the SM \ttbar{} background: 
the initial- and final-state radiation (ISR/FSR) systematic uncertainty, the fragmentation and parton shower (PS) systematic uncertainty based on a comparison of {\sc Pythia} and {\sc Herwig}, 
and the generator systematic uncertainty based on a comparison of {\sc MC@NLO} and {\sc Powheg}. 
These uncertainties individually lead to 3$-$6\% variations in the total background yield. In addition, the \ttbar{} normalization is affected by the uncertainty in the theoretical prediction of the \ttbar{} production cross section.

The impact of the luminosity uncertainty (nominally 3.7\%) is reduced for the background, compared with the signal, because two of the backgrounds, $W$+jets and multijets, are determined from data.

The normalization of the $W$+jets background using the charge
asymmetry method is described in Section~\ref{sec:backgroundwjets}. 
The statistical uncertainty on the $W$+jets yield amounts to less than 10\%. 
A systematic uncertainty of 14\% is assigned to account for
systematic uncertainties in the background subtraction 
(normalization of the subtracted backgrounds is varied by 100\%),  
the PDFs, the jet energy scale uncertainty and the $W$+jets modelling. 
The uncertainty in normalization and shape 
of the $W$+jets contribution due to the extrapolation from the
control region to the signal region is accounted for in
the jet scale and resolution uncertainty
in Table~\ref{tab:systematics}. If this contribution is added, the 
total uncertainty on the $W$+jets yield amounts to approximately 35\%.
Repeating the normalization 
procedure on several $W$+jets validation regions we find that the result is 
always consistent within the assigned systematic uncertainty.
The shape of the $W$+jets contribution to the $m_{t \bar{t}}$ distribution 
is estimated from simulation. The shape uncertainty includes the
impact of the systematic uncertainties on jets and a $W$+jets modelling 
uncertainty obtained by varying a number of parameters in 
{\sc Alpgen}, such as the factorization scale and the 
scale governing the value of the strong coupling constant $\alpha_s$ used 
in the parton splittings.

A constant normalization uncertainty of 50\% is applied to the data-driven 
multijet background estimation, as well as a shape 
uncertainty derived by comparing two different 
loose lepton selection criteria.

The uncertainties related to charged lepton reconstruction are labelled as ``Electron'' and ``Muon'' in the table. 
The \MET{} uncertainty is negligible 
and is not listed in Table~\ref{tab:systematics}. Among the uncertainties on
reconstructed objects, those related to jets are the most important. 
The uncertainty on the scales for the jet energy and mass 
measurements is estimated from a combination 
of in situ measurements, test beam data and 
Monte Carlo studies~\cite{Aad:2011he}. The energy scale uncertainty
for anti-$k_t$ jets with $R=0.4$ is less than 3\% in the range of energies
relevant for this search. The energy scale for $R=1.0$ jets is only 
slightly larger, while the uncertainty on the jet mass scale (JMS) is 
approximately 4$-$5\%~\cite{Aad:2012ef}. Energy deposits due to additional 
proton-proton interactions (pile-up) have a strong impact on the measured 
mass of fat jets. An additional 1\% uncertainty on the JMS 
accounts for imperfections in the pile-up model and non-closure of the 
Monte Carlo reweighting procedure.
The effects of these uncertainties on the event yields and sensitivity are shown in Table~\ref{tab:systematics}.

\section{Comparison of data to the Standard Model prediction}
\label{sec:datamc}
The selection described in Section~\ref{sec:selection} yields a total of 1837 data events, in agreement with the Standard Model  prediction ($1840 \pm 130$ events). 
The background expectation is broken down by source in Table~\ref{tab:datamcyield}, separately for $e$+jets and $\mu$+jets channels. 

\begin{table}[tbh!]
\begin{center}
\caption{Selected data events and expected background yields after the full selection. The uncertainties on the normalization of the expected background yield are also listed.
}
\vspace{2mm}
\begin{tabular}{lr@{$~\pm~$}lr@{$~\pm~$}lr@{$~\pm~$}l}
\hline
Type          & \multicolumn{2}{c}{$e$+jets}  & \multicolumn{2}{c}{$\mu$+jets} & \multicolumn{2}{c}{Sum} \\  \hline 
\ttbar{}   &510 & 40 & 620 & 50 & 1130 & 90 \\
$W$+jets   &202 & 34 & 300 & 50 &  500 & 80\\
Multijets  & 45 & 23 &  30 & 15 &   75 & 38\\
$Z$+jets   & 41 & 20 &  34 & 16 &   75 & 36\\
Single top & 21 & 2  &  27 & 3  &   48 &  5\\
Dibosons   & 3.4& 0.2& 4.5 & 0.2&   7.9& 0.4\\
Total      &830 & 60 &1010 & 70 & 1840 & 130\\
\hline
Data & \multicolumn{2}{c}{803} & \multicolumn{2}{c}{1034} & \multicolumn{2}{c}{1837} \\
\hline
\end{tabular}
\label{tab:datamcyield}
\end{center}
\end{table}

\begin{figure}[htbp!]
\begin{center}
\subfigure[]{
  \label{fig:datamc_akt10pt}
  \includegraphics[width=0.7\textwidth]{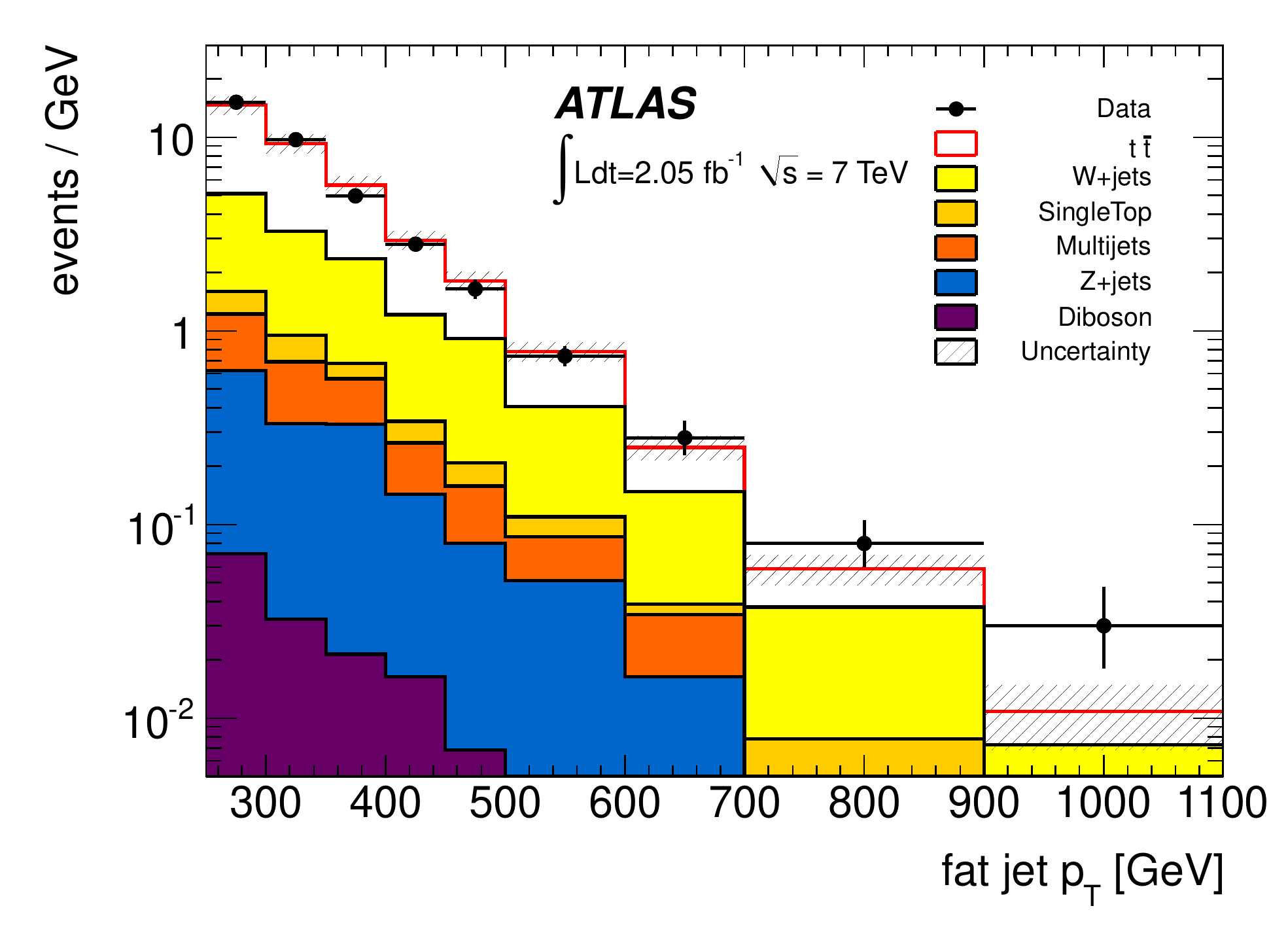}
}
\subfigure[]{
  \label{fig:datamc_akt10masslog}
  \includegraphics[width=0.7\textwidth]{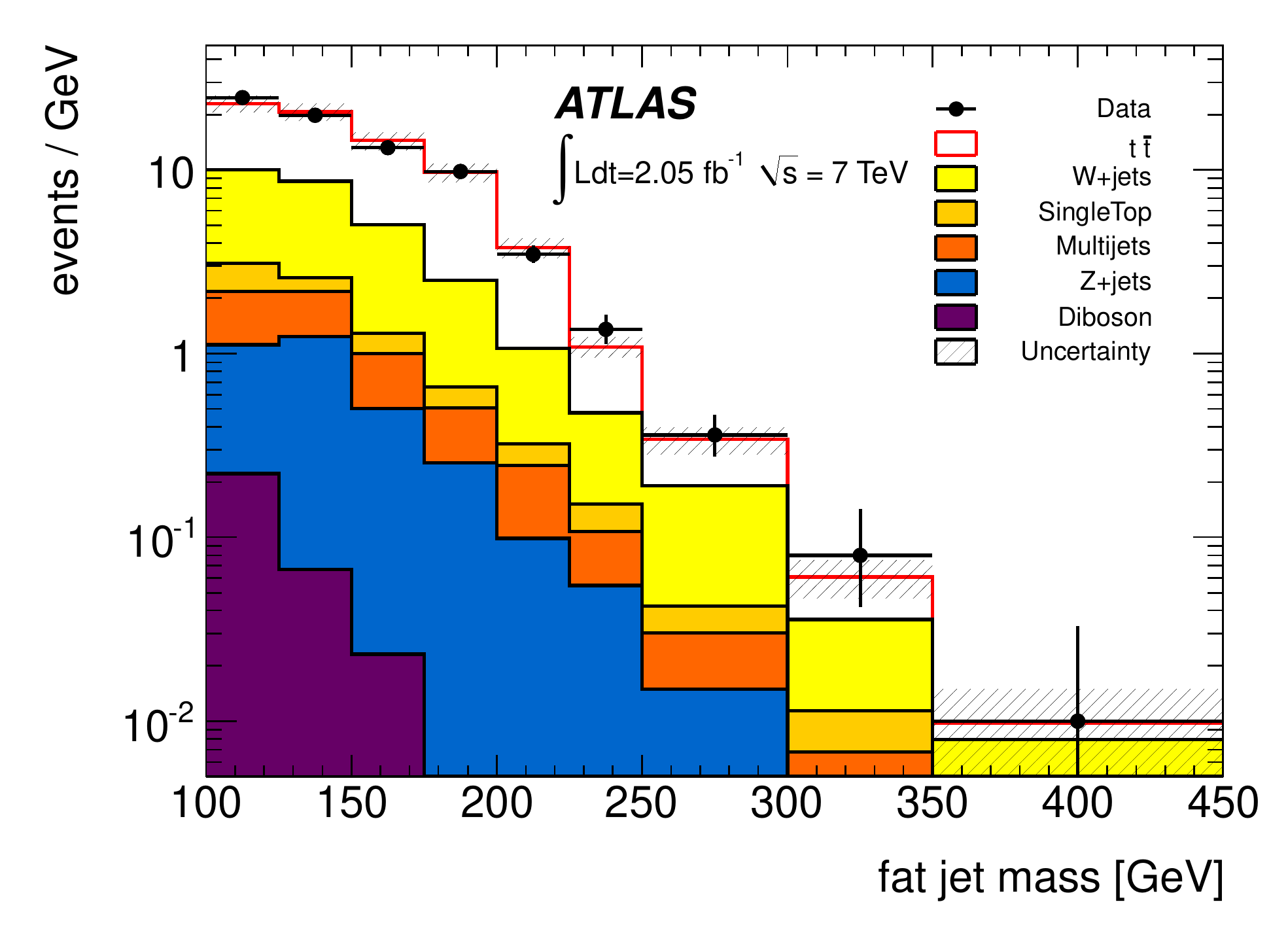}
}
\caption{Comparison of the data and the Standard Model prediction for two kinematic distributions: (a) transverse momentum and (b) jet mass of the fat $R= 1.0$ jets selected as the hadronically decaying top quark candidate. The $e$+jets and $\mu$+jets channels are combined. The shaded band indicates the normalization uncertainty on the Standard Model prediction, but does not include the shape uncertainty or the impact of uncertainties on reconstructed objects.}
\label{fig:datamc}
\end{center}
\end{figure}

\begin{figure}[htbp]
\begin{center}
  \includegraphics[width=0.7\textwidth]{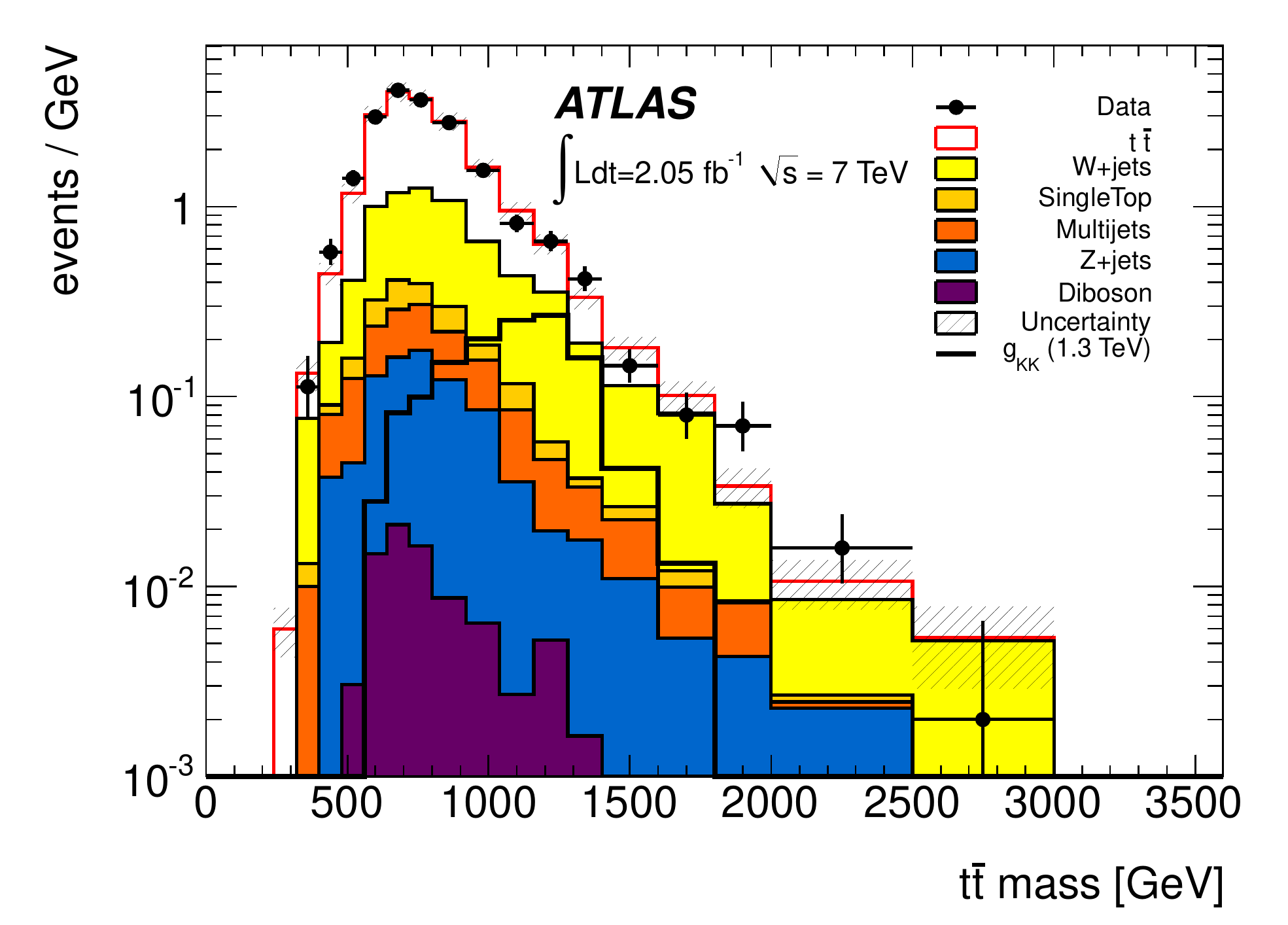}
\caption{Reconstructed invariant mass distribution of the $t \bar{t}$ candidates after signal selection. The $e$+jets and $\mu$+jets channels are combined. The shaded band indicates the uncertainty in the normalization of the Standard Model prediction, but does not include the shape uncertainty or the impact of uncertainties on reconstructed objects. The variable bin size is chosen to match the mass resolution for a resonant signal.}
\label{fig:ttbar_mass_spectra}
\end{center}
\end{figure}

The distributions of two key observables, the transverse momentum and the invariant mass of the fat $R= 1.0$ jet that is selected as the hadronically decaying top quark candidate, are compared with the Standard Model predictions in Figure~\ref{fig:datamc}. 
The data are found to agree with the expectation within the error band that indicates the normalization uncertainty.
The reconstructed \ttbar{} mass spectrum is presented in Figure~\ref{fig:ttbar_mass_spectra}.

\section{Interpretation}
\label{sec:limits}

The compatibility of the data with the SM-only (null) hypothesis is evaluated with the BumpHunter code~\cite{Choudalakis:2011qn}, a tool that searches for local data excesses or deficits of varying width compared to the expected background. 
The most significant excess is found in the \ttbar{} mass region between 1.8 and 2.5~\tev. It is most pronounced in the electron channel. 
When the systematic uncertainties are accounted for, the $p$-value or probability that the observed excess is found under the assumption that the null hypothesis (Standard Model) is true, is 0.08 \mbox{(1.4$\sigma$)}, including the look-elsewhere effect, evaluated over the full mass range. 
No other deviations with respect to the Standard Model prediction with a significance beyond 1$\sigma$ are found. 

We set 95\% CL upper limits on the production cross section times branching fraction of new massive states using Bayesian techniques~\cite{topstatistics}. 
The prior probability distribution used in this method, which is flat in the cross section, is a good approximation of the reference prior~\cite{Casadei_prior}, and the likelihood is calculated using a Poisson function. 
The systematic uncertainties described in Section~\ref{sec:systematics} are found to have a significant impact on the sensitivity: in the 1.0$-$1.5~\tev{} mass range the limit on the rate including all systematic uncertainties is typically a factor two weaker than the limit that would be derived with statistical uncertainties only. 
The systematic uncertainties are accounted for by assuming they are normally distributed and convolving a Gaussian with the posterior probability distribution for each one. 
They are only weakly constrained by the data. 
Two exceptions are the jet energy scale and the \ttbar{} generator dependence, which are constrained to about a half or a third, respectively, of their prior uncertainty. 
The result has been cross-checked with the so-called $CL_s$ method~\cite{Junk:1999kv,Read:2002hq} and is in good agreement with it. 
Not allowing the data to constrain the systematic uncertainties in the $CL_s$ method reduces the sensitivity on the signal cross section by no more than 20\%, which is less than the expected 1$\sigma$ variation.

\begin{figure}[tbp]
\begin{center}
\subfigure[]{
\label{F:limits_Zprime}
\includegraphics[width=0.6\textwidth]{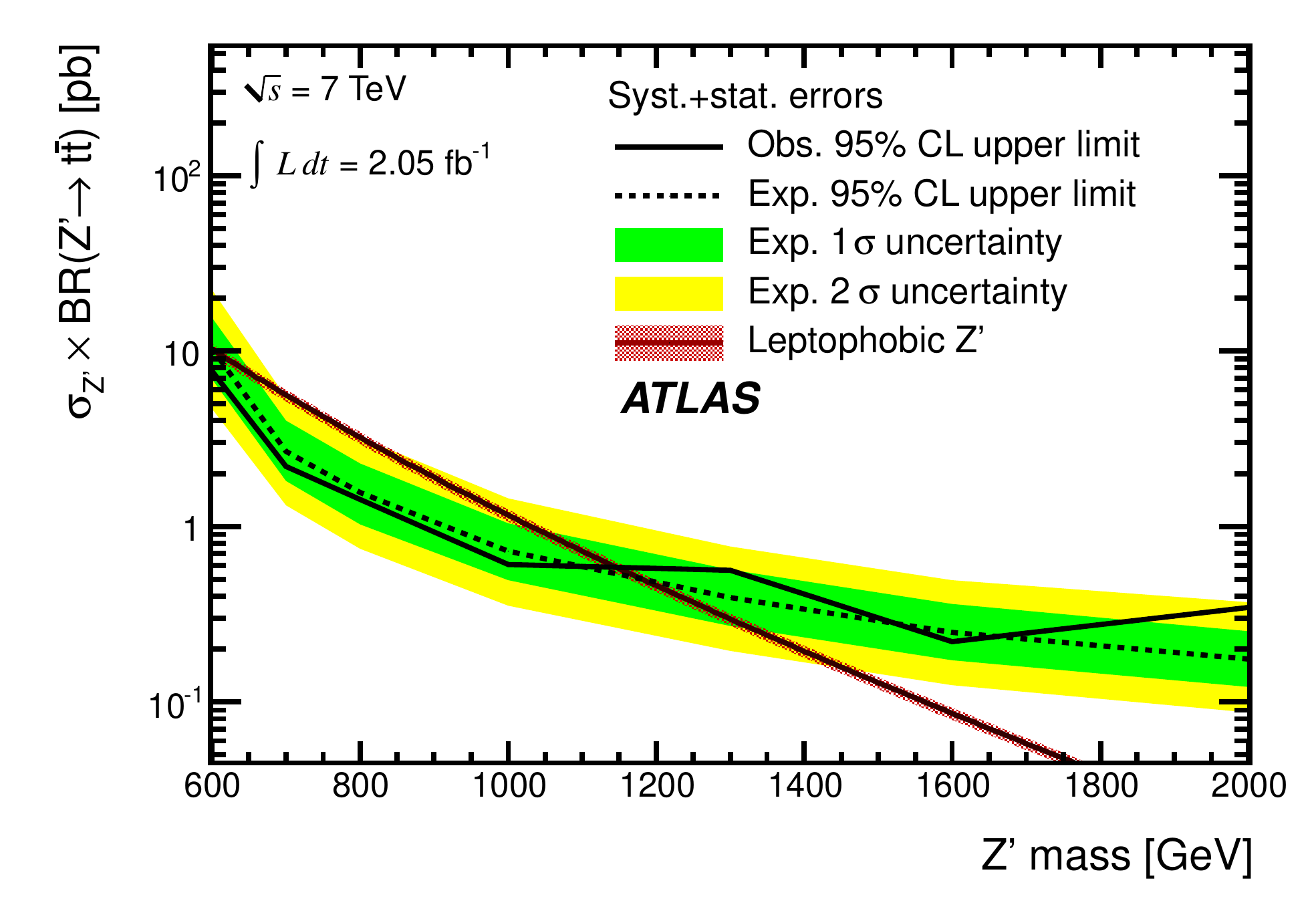} }
\subfigure[]{
\label{F:limits_kkg}
\includegraphics[width=0.6\textwidth]{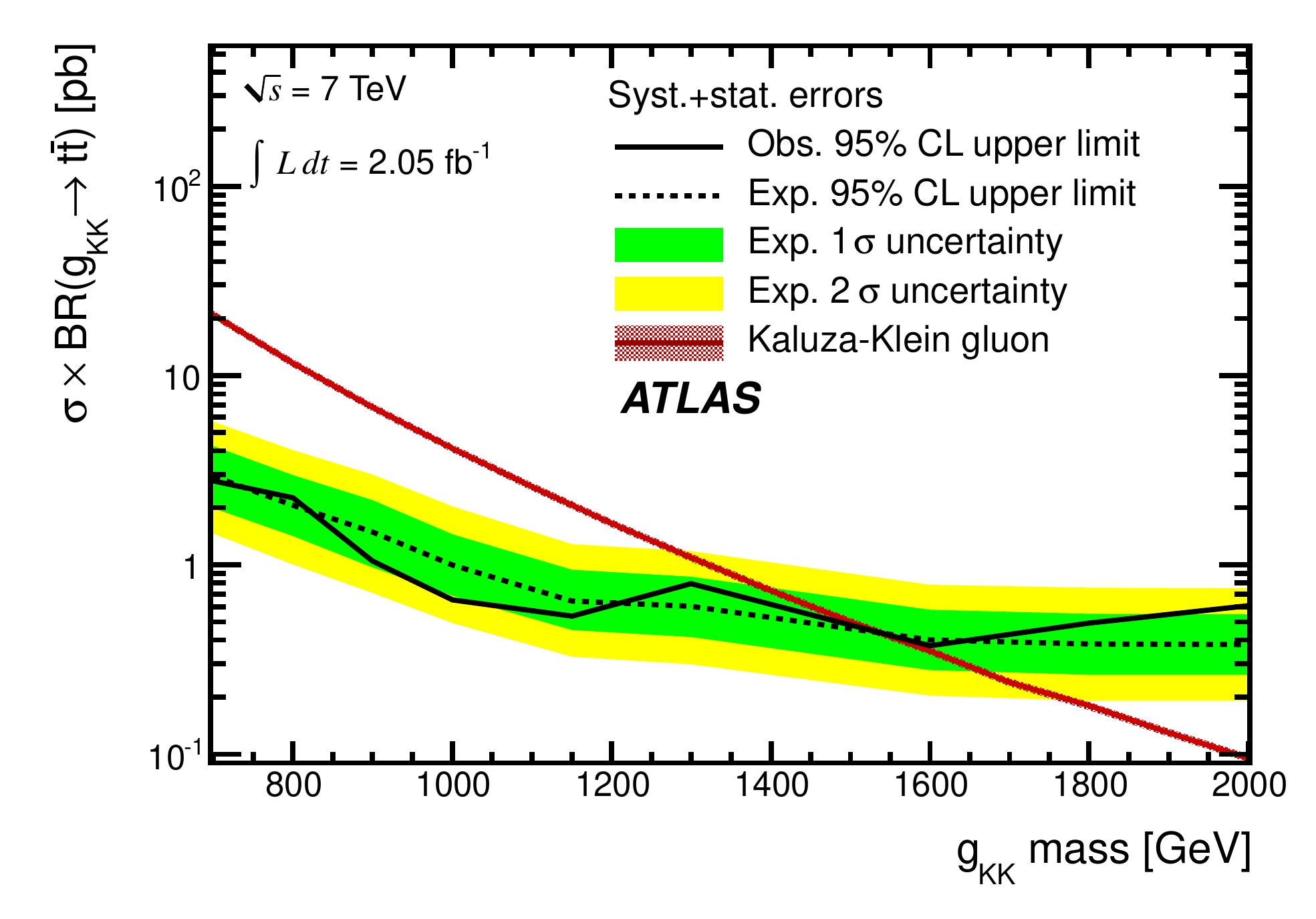} }\\
\caption[Upper production cross section limits, systematic and statistical uncertainties]{%
Expected (dashed line) and observed (solid line) upper limits on the production cross section times the \ttbar{} branching fraction of (a) \Zprime{} and (b) Kaluza--Klein gluons. The dark (green) and light (yellow) bands show the range in which the limit is expected to lie in 68\% and 95\% of pseudo-experiments,
respectively, and the smooth solid (red) lines correspond to the predicted production cross section times branching fraction for the \Zprime\ (a) and Randall--Sundrum (b) models. 
The band around the signal cross section curve is based on the effect of the PDF uncertainty on the prediction. \label{fig:limit_Zprime_kkg}}
\end{center}
\end{figure}

The resulting limits for narrow \Zprime{} resonances and broad coloured resonances are presented in Figure~\ref{fig:limit_Zprime_kkg} and Table~\ref{tab:limits}. 
Upper limits on the production cross section times branching fraction of a narrow \Zprime{} resonance range from approximately 8~pb for a \Zprime{} mass of 0.6~\tev{} to 610~fb at 1~\tev{} and 220~fb at 1.6~\tev{}, in good agreement with the expected limits. 
These lead to exclusion of the mass range between 0.6~\tev{} and \upperZprimemasslimit~\tev{} for the leptophobic topcolor \Zprime{} model considered here. 
The upper value of the excluded mass range agrees with the expected sensitivity within the estimated precision. The observed upper cross section limits for a \Zprime\ and a Kaluza--Klein gluon at 2 \TeV\ are about 1.5$\sigma$ higher than the expected value, reflecting the small data excess at 1.8-2.5 \TeV, seen in Figure~\ref{fig:ttbar_mass_spectra}. 

The observed limits on the broad ($\Gamma/m =15.3$\%) Kaluza--Klein gluon are slightly weaker than those on the \Zprime{} boson: 650~fb at 1~\tev{} and 370~fb at 1.6~\tev{}. 
The impact of the width is observed also in the expected limits and is most pronounced for large resonance masses: at 1~\tev{} the expected limit on the \Zprime{} boson is stronger by approximately 30\%, but the ratio of the \Zprime{} limit to the KK gluon limit reaches two at 2~\tev{}. 
The KK gluon model of Lillie et al.~\cite{Lillie:2007yh} is excluded for a resonance mass below \upperKKgmasslimit~\tev{}, again in good agreement with the expectation.

\begin{table}[tbh!]
\begin{center}
\caption{Observed and expected upper limits on the production cross section times branching fraction for $\Zprime \rightarrow \ttbar$ and $g_{KK} \rightarrow \ttbar$ respectively, including systematic and statistical uncertainties. The expected limit $\pm 1 \sigma$ variation is also given.
}
\vspace{2mm}
\begin{tabular}{rr@{.}lr@{.}lr@{.}lr@{.}l}
\hline
\multicolumn{9}{c}{$\Zprime \rightarrow \ttbar$ limits} \\
\Zprime\ Mass [GeV] &  \multicolumn{2}{c}{Observed [pb]}  &  \multicolumn{2}{c}{Expected [pb]}  &  \multicolumn{2}{c}{$-$1$\sigma$ [pb]}  &  \multicolumn{2}{c}{$+$1$\sigma$[pb]} \\
\hline
 600  &  \hspace{2em}7&7   &  \hspace{2em}10&4    & \hspace{1ex}7&0   &  \hspace{1ex}15&6 \\
 700  &  2&2   &   2&7    & 1&8   &   4&0 \\
 800  &  1&4   &   1&6    & 1&0   &   2&3 \\
1000  &  0&61  &   0&72   & 0&49  &   1&0 \\
1300  &  0&56  &   0&39   & 0&27  &   0&57 \\
1600  &  0&22  &   0&25   & 0&17  &   0&36 \\
2000  &  0&34  &   0&18   & 0&12  &   0&25 \\
3000  &  0&27  &   0&27   & 0&19  &   0&41 \\
\hline 
\multicolumn{9}{c}{$g_{\mathrm{KK}} \rightarrow \ttbar$ limits} \\
$g_{\mathrm{KK}}$ Mass [GeV]  &  \multicolumn{2}{c}{Observed [pb]}  &  \multicolumn{2}{c}{Expected [pb]}  &  \multicolumn{2}{c}{$-$1$\sigma$ [pb]}  &  \multicolumn{2}{c}{$+$1$\sigma$[pb]} \\
\hline
 700  &  2&8   &  2&9   &  2&0   &  4&2 \\
 800  &  2&3   &  2&1   &  1&4   &  3&0 \\
 900  &  1&0   &  1&5   &  0&97  &  2&2 \\
1000  &  0&65  &  0&99  &  0&69  &  1&4 \\
1150  &  0&53  &  0&64  &  0&45  &  0&94 \\
1300  &  0&80  &  0&60  &  0&42  &  0&87 \\
1600  &  0&37  &  0&40  &  0&28  &  0&58 \\
1800  &  0&49  &  0&38  &  0&26  &  0&55 \\
2000  &  0&61  &  0&38  &  0&26  &  0&55 \\
\hline
\end{tabular}
\label{tab:limits}
\end{center}
\end{table}

To establish the potential of a selection and reconstruction scheme specifically designed for \ttbar{} events with highly boosted top quarks, it is instructive to compare the current result to a previous analysis of the same data set~\cite{resolved}. 
We observe that the current search leads to significantly better limits in the 1$-$2~\tev{} range, where the expected limit on the production cross section times branching fraction for a narrow resonance is improved by a factor 1.5 to 2.
 It should be noted that the current result represents a partial implementation of the algorithm designed for highly boosted top quarks~\cite{ATL-PHYS-PUB-2010-008}, especially in the isolation requirement in the lepton selection.

\section{Summary}

We present results of a search for resonances in the $t \bar{t}$ mass spectrum in a 2.05~\ifb{} data set collected with the ATLAS detector during the 2011 proton-proton runs of the Large Hadron Collider  at a centre-of-mass energy of 7~\tev. 
The analysis focuses on the lepton+jets final state obtained when one $W$ boson decays to a charged lepton and a neutrino, and the other decays to a quark and an anti-quark pair. 
The selection and reconstruction are specifically designed for the collimated topology that arises from the decay of boosted top quarks. 
The hadronically decaying top quark candidate is identified as a single jet with radius parameter $R=1.0$ that is required to have significant substructure, as measured by the jet mass and $k_t$ splitting scale.

The reconstructed $t \bar{t}$ mass spectrum is compared with a template for the Standard Model prediction constructed using a combination of Monte Carlo simulations and measurements using control samples. 
The data are found to be compatible with the SM within uncertainties. 
Upper limits at 95\% CL on the production cross section times branching ratio of the narrow \Zprime{} resonance range from approximately 8 pb for a mass of 600~\gev{} to 220 fb for a mass of 1.6~\tev{}, 
in good agreement with the expected limits. 
These lead to exclusion of the mass range between 600~\gev{} and \upperZprimemasslimit~\tev{} for the leptophobic topcolor \Zprime{} model considered here. 
Slightly weaker limits are derived on a broad resonance ($\Gamma/m = 15.3$\%). 
The KK gluon model of Lillie et al.~\cite{Lillie:2007yh} is excluded for a resonance mass below \upperKKgmasslimit~\tev{}. 
These resonance searches reach the sensitivity in the high-mass region that they can verify or exclude recent proposals such as that of Djouadi et al.~\cite{Djouadi:2011aj}. 
The sensitivity of this search geared to highly boosted top quarks is significantly enhanced in the 1$-$2~\tev{} region with respect to a previously published search using the same data set~\cite{resolved}. 
The limit on the KK gluon is the most stringent limit on this model to date.

\section{Acknowledgements}

We thank CERN for the very successful operation of the LHC, as well as the
support staff from our institutions without whom ATLAS could not be
operated efficiently.

We acknowledge the support of ANPCyT, Argentina; YerPhI, Armenia; ARC,
Australia; BMWF, Austria; ANAS, Azerbaijan; SSTC, Belarus; CNPq and FAPESP,
Brazil; NSERC, NRC and CFI, Canada; CERN; CONICYT, Chile; CAS, MOST and NSFC,
China; COLCIENCIAS, Colombia; MSMT CR, MPO CR and VSC CR, Czech Republic;
DNRF, DNSRC and Lundbeck Foundation, Denmark; EPLANET and ERC, European Union;
IN2P3-CNRS, CEA-DSM/IRFU, France; GNAS, Georgia; BMBF, DFG, HGF, MPG and AvH
Foundation, Germany; GSRT, Greece; ISF, MINERVA, GIF, DIP and Benoziyo Center,
Israel; INFN, Italy; MEXT and JSPS, Japan; CNRST, Morocco; FOM and NWO,
Netherlands; RCN, Norway; MNiSW, Poland; GRICES and FCT, Portugal; MERYS
(MECTS), Romania; MES of Russia and ROSATOM, Russian Federation; JINR; MSTD,
Serbia; MSSR, Slovakia; ARRS and MVZT, Slovenia; DST/NRF, South Africa;
MICINN, Spain; SRC and Wallenberg Foundation, Sweden; SER, SNSF and Cantons of
Bern and Geneva, Switzerland; NSC, Taiwan; TAEK, Turkey; STFC, the Royal
Society and Leverhulme Trust, United Kingdom; DOE and NSF, United States of
America.

The crucial computing support from all WLCG partners is acknowledged
gratefully, in particular from CERN and the ATLAS Tier-1 facilities at
TRIUMF (Canada), NDGF (Denmark, Norway, Sweden), CC-IN2P3 (France),
KIT/GridKA (Germany), INFN-CNAF (Italy), NL-T1 (Netherlands), PIC (Spain),
ASGC (Taiwan), RAL (UK) and BNL (USA) and in the Tier-2 facilities
worldwide.


\bibliographystyle{JHEP}
\bibliography{./boosted,./PHYSICS}

\clearpage
\begin{flushleft}
{\Large The ATLAS Collaboration}

\bigskip

G.~Aad$^{\rm 47}$,
T.~Abajyan$^{\rm 20}$,
B.~Abbott$^{\rm 110}$,
J.~Abdallah$^{\rm 11}$,
S.~Abdel~Khalek$^{\rm 114}$,
A.A.~Abdelalim$^{\rm 48}$,
O.~Abdinov$^{\rm 10}$,
R.~Aben$^{\rm 104}$,
B.~Abi$^{\rm 111}$,
M.~Abolins$^{\rm 87}$
% ...
\bigskip

% ...
$^{10}$ Institute of Physics, Azerbaijan Academy of Sciences, Baku, Azerbaijan\\
$^{11}$ Institut de F\'{i}sica d'Altes Energies and Departament de F\'{i}sica de la Universitat Aut\`{o}noma de Barcelona and ICREA, Barcelona, Spain\\
% ...
$^{20}$ Physikalisches Institut, University of Bonn, Bonn, Germany\\
% ...
$^{47}$ Fakult\"{a}t f\"{u}r Mathematik und Physik, Albert-Ludwigs-Universit\"{a}t, Freiburg, Germany\\
$^{48}$ Section de Physique, Universit\'{e} de Gen\`{e}ve, Geneva, Switzerland\\
% ...
$^{87}$ Department of Physics and Astronomy, Michigan State University, East Lansing MI, United States of America\\
% ...
$^{104}$ Nikhef National Institute for Subatomic Physics and University of Amsterdam, Amsterdam, Netherlands\\
% ...
$^{110}$ Homer L. Dodge Department of Physics and Astronomy, University of Oklahoma, Norman OK, United States of America\\
$^{111}$ Department of Physics, Oklahoma State University, Stillwater OK, United States of America\\
% ...
$^{114}$ LAL, Universit\'{e} Paris-Sud and CNRS/IN2P3, Orsay, France\\
% ...
\end{flushleft}


\end{document}